\documentclass[aps,pra,twocolumn,showpacs,amsmath,amssymb,superscriptaddress,bibliography]{revtex4-2}
\usepackage{graphicx}    % needed for figures
\usepackage{bm}             % for math
\usepackage{subfigure}    % for subfigure
\usepackage{braket}
\usepackage{gensymb}
\usepackage{hyperref} % For clickable links
\usepackage[toc,page]{appendix}
\graphicspath{{Figures/}{./}}
\begin{document}
% \message{The column width is: \the\columnwidth}
% The column width is: \the\columnwidth

\title{Cooling trapped ions with phonon rapid adiabatic passage}
\author{M. I. Fabrikant}
\address{Quantinuum, 303 S Technology Ct, Broomfield, CO 80021, USA}
\author{P. Lauria}
\address{Quantinuum, 303 S Technology Ct, Broomfield, CO 80021, USA}
\author{I. S. Madjarov}
\address{Quantinuum, 303 S Technology Ct, Broomfield, CO 80021, USA}
\author{W. C. Burton}
\address{Quantinuum, 303 S Technology Ct, Broomfield, CO 80021, USA}
\author{R. T. Sutherland}
\email{robert.sutherland@quantinuum.com}
\address{Quantinuum, 303 S Technology Ct, Broomfield, CO 80021, USA}
\address{Department of Electrical and Computer Engineering, University of Texas at San Antonio, San Antonio, Texas 78249, USA}

\date{\today}

\begin{abstract}
In recent demonstrations of the quantum charge-coupled device (QCCD) computer architecture, circuit times are dominated by cooling. Some motional modes of multi-ion crystals take orders-of-magnitude longer to cool than others because of low coolant ion participation. Here we demonstrate a new technique, which we call phonon rapid adiabatic passage (phrap), that avoids this issue by coherently exchanging the thermal populations of selected modes on timescales short compared to direct cooling. Analogous to adiabatic rapid passage, we quasi-statically couple these slow-cooling modes with fast-cooling ones using DC electric fields. When the crystal is then adiabatically ramped through the resultant avoided crossing, nearly-complete phonon population exchange results. We demonstrate this on two-ion crystals, and show the indirect ground-state cooling of all radial modes\textemdash achieving an order of magnitude speedup compared to direct cooling. We also show the technique's insensitivity to trap potential and control field fluctuations, and find that it still achieves sub-quanta temperatures starting as high as $\bar{n}\sim 200$.
\end{abstract}
\maketitle

\section{Introduction}\label{sec:intro}
Trapped ions are a useful tool for quantum information and metrology \cite{wineland_1998,blatt_2008,timoney_2011,nielsen_2010,pino_2021,moses_2023}, due, in part, to the fact that we can precisely manipulate their internal states \cite{cirac_1995,monroe_1995,molmer_1999,molmer_2000,mintert_2001,leibfried_2003,harty_2014,ballance_2016,gaebler_2016,weidt_2016,sutherland_2019,sutherland_2020,srinivas_2021,clark_2021,sutherland_2024}. Generally, we increase this control by cooling relevant motional modes to near their ground state \cite{wineland_1998}, which can improve the fidelity of two-qubit gates \cite{ballance_2016,gaebler_2016,harty_2016,webb_2018,li_2020,srinivas_2021,sutherland_2022_1} and the precision of atomic clocks \cite{ludlow_2015,chou_2010,barwood_2014,huntemann_2016,chen_2017,kim_2023}\textemdash often using a sympathetic coolant ion of a different species than the qubit/clock ion, since this can help prevent crosstalk and other technical issues \cite{kielpinski_2000,kielpinski_2002,blinov_2002,pino_2021,moses_2023}. Of particular interest are computers based on the quantum charge coupled device (QCCD) architecture \cite{kielpinski_2002}, which use surface Paul traps \cite{paul_1990,chiaverini_2005,seidelin_2006,seidelin_2006, leibrandt_2009, schulz_2008} to confine and shuttle qubit/coolant crystals to and from designated trap zones \cite{pino_2021,moses_2023}. Paul traps confine each ion with a ponderomotive `pseudopotential' transverse to the trap axis, with a depth that scales with mass like $1/m$. In order to prevent the decoupling of radial modes into `slow-cooling' mostly-qubit modes and `fast-cooling' mostly-coolant modes \cite{wubbena_2012,sosnova_2021}, the ion mass ratio must be kept near unity. For example, switching from $^{27}$Al$^{+}$-$^{9}$Be$^{+}$ to $^{27}$Al$^{+}$-$^{25}$Mg$^{+}$ enabled more effective cooling in clocks \cite{rosenband_2007,rosenband_2008,chen_2017}, decreasing errors from thermal motion by over an order-of-magnitude. Similarly, many gating schemes directly couple the qubits to radial modes \cite{zhu_2006,kim_2009,serafini_2009,ospelkaus_2011,green_2015,harty_2016,debnath_2016,bermudez_2017,srinivas_2021,zarantonello_2019,jeon_2023}, making entropy removal from these modes especially important. \\ 

\begin{figure*}[t]
\includegraphics[width=1\textwidth]{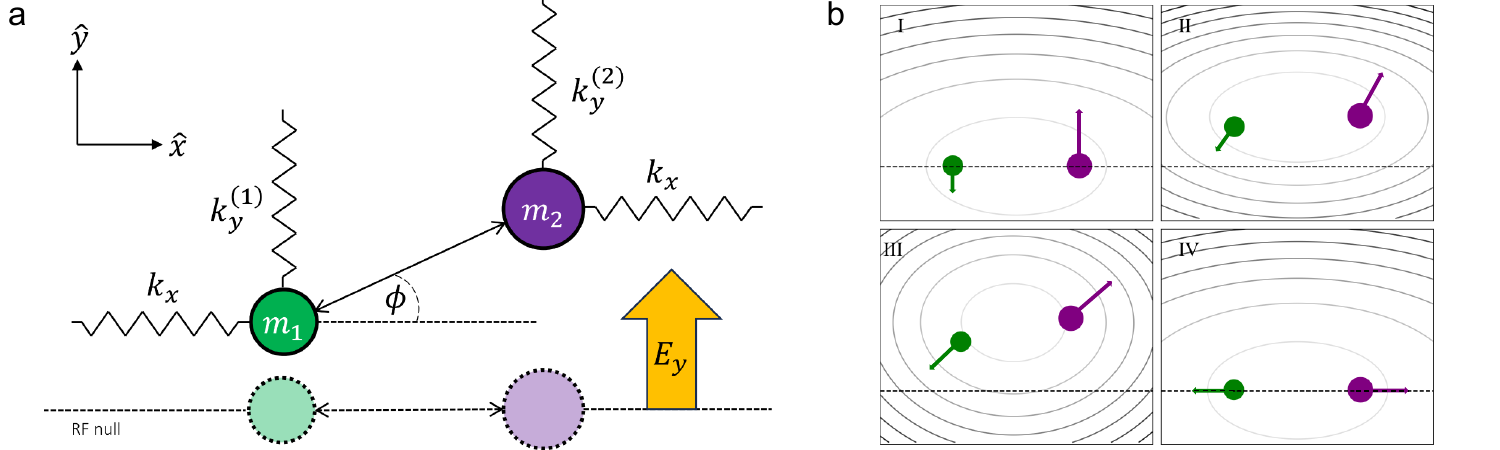}
\caption[Fig1]{\textbf{Phrap technique overview.} (a) A trapped crystal of two ions with masses $m_1 \neq m_2$ is displaced off the trap's RF null (lower dashed line) in the radial $y$-direction by a spatially homogeneous electric field $E_y$. Because of the mass-dependence of the RF pseudopotential, the ions experience different radial confinements $k_y^{(1)} \neq k_y^{(2)}$ and are therefore displaced off the null by different amounts. This creates an angle $\phi$ between the principal trapping axes of each ion and the axis of the Coulomb interaction between them, coupling axial ($x$-direction) and radial ($y$-direction) modes. (b) Snapshots of the physical positions and displacement vectors of ions undergoing phrap. (I) Both ions begin on the RF null with displacement vectors consistent with the to-be-phraped eigenmode. (II) The coupling shim is added. (III) The axial trap frequency is ramped past the point of axial-radial mode degeneracy. (IV) The coupling shim is turned off and the axial frequency quickly relaxes to the original value. At the end of the relaxation, the ion motion has swapped from the initial mode to the target mode.}
\label{fig:fig_1}
\end{figure*}

Some schemes have attempted to cool these modes indirectly by first cooling a fast-cooling mode, coherently exchanging its thermal population with a slow-cooling mode, and then repeating the cooling step\textemdash essentially turning the first mode into a heat sink. This general idea is promising, but developing methods that straightforwardly integrate into QCCD architectures is challenging. One way to couple/exchange modes is to use their internal states as a bus \cite{king_2021,sutherland_2021_cvqc}. Unfortunately, this necessitates entangling the ions with the mode we intend to cool, indicating that, to cool a mostly-qubit mode, you need to entangle your qubit ion to its motion\textemdash exposing the qubit to decoherence. It is also possible to couple modes via the Kerr-like terms in the Coulomb potential \cite{marquet_2003,ding_2017_counting,ding_2017}, but this interaction is limited in strength and difficult to control. Alternatively, modes may be electronically coupled by modulating the harmonic potential at the difference frequency between two modes using only the electrodes \cite{gorman_2014,hou_2022,hou_2023}. This technique can be described by an effective phonon exchange (beam-splitter) Hamiltonian \footnote{The operator component of $\hat{H}_{\text{p}}$ belongs to a group that follows SU(2) algebra, meaning, in the Heisenberg picture, they transform in the same was that Pauli operators do.}, driving a `Rabi flop' that fully exchanges the modes' populations at specified times \cite{gorman_2014}. However, this scheme has several drawbacks. First, motional mode frequencies are usually $\sim \text{MHz}$, well outside the typical electrode filter bandwidth, limiting the achievable drive strengths. It is also relatively sensitive to the control field's amplitude and frequency, requiring careful local calibration to compensate for the variations in trap potential which are unavoidable in any large-scale system. Yet scalable architectures must rely on broadcasted waveforms to budget the total signal count \cite{malinowski_2023}, and it is not possible to optimize a single broadcasted waveform for every crystal across the trap. 

In this work, we present a new method for transferring thermal population between the motional modes of a crystal, inspired by rapid adiabatic passage ~\cite{malinovsky_2001}. This technique, which we term phonon rapid adiabatic passage (phrap), uses quasi-static control signals and is less sensitive to parameter fluctuations than the Rabi-flop schemes discussed above \footnote{Since we are interested in cooling here, and thus ambivalent to motional phase coherence, it is reasonable to expect a scheme based on rapid adiabatic passage for transferring thermal population between motional modes to be more robust than phonon pumping, which we explore below.}. To implement phrap, we manipulate the control electrodes of our trap to bring the frequency of a cold motional mode ($\bar{n}\sim 0$) and a hot motional mode ($\bar{n}\sim 10$) close to degeneracy. While doing this, we use an electric field to push the ions off the RF null of the trap. Because of the $1/m$ dependence of the pseudopotential, this field pushes the differently-massed ions different distances, giving the Coulomb force a mutual projection onto the axial and radial directions (see Fig.~\ref{fig:fig_1}). This couples the radial and axial modes, opening an avoided crossing we use to drive the phonon exchange. We ramp the control electrodes to pass through the avoided crossing at a rate that is slow relative to the crossing's splitting, maintaining each mode's state while swapping their characters (see Fig.~\ref{fig:example_phrap_abc}). We then turn off the coupling and diabatically return to the original well through the now-closed mode crossing, completing the mode exchange. There is evidence that similar transitions can occur, inadvertently, in certain transport operations, limiting performance by allowing heat to leak between axial and radial modes \cite{lancellotti_2023}. \\

Specifically, in Sec.~\ref{sec:theory} we present the theory of phrap, analyzing the scheme for $2$-modes and then generalizing to $N$-modes. We then discuss how radial electric fields can induce the type of mode-mode coupling needed for phrap\textemdash only in mixed-species crystals. We also point out what are likely to be the biggest control error mechanisms. Then, in Sec.~\ref{sec:experiments}, we experimentally demonstrate phrap on Ba-Yb crystals. First, we show how radial electric fields can couple two modes, showing the associated avoided crossing. Then, we demonstrate the full-transfer of population between two targeted modes. We also show the scheme's robustness to control parameters, confirming that it remains efficient for a wide range of sequence durations, trap confinements, and mode temperatures. We then demonstrate two ways to achieve all-mode-cooling with direct cooling only on a subset of modes. In Sec.~\ref{sec:gen_all_mode_cooling}, we discuss a scheme that allows all-mode-cooling, given direct cooling on any crystal mode. Finally, in Sec.~\ref{sec:conclusion}, we offer a conclusion.

\section{Theory}\label{sec:theory}

We consider a two ion crystal in an `uncoupled' Paul trap, where each ion $j$ experiences the potential: 
\begin{eqnarray}\label{eq:paul_trap_unrot}
    V_{\text{t},j} = V_{xx,j}x^{2}_{j} + V_{yy,j}y^{2}_{j} + V_{zz,j}z^{2}_{j},
\end{eqnarray}
where we take $x$ to be the crystal axis, while $r\in\{y,z\}$ are the two radial directions. The curvatures are:
\begin{eqnarray}\label{eq:potentials}
    V_{xx,j} &=& V_{\text{DC}}, V_{rr,j} = \frac{V_{\text{RF}}^{2}}{\Omega_{\text{RF}}^{2}m_{j}} - c_{r}V_{\text{DC}},
\end{eqnarray}
where $V_{\text{DC}}$ is the DC curvature that provides axial confinement, $c_{y}+c_{z}=1$ are the corespondent fractional decreases in radial curvature, and $m_{j}$ is the mass of ion $j$. The terms $V_{\text{RF}}$ and $\Omega_{\text{RF}}$ are the strength and frequency of the RF drive that creates the pseudopotential. Importantly here, the pseudopotential term scales inversely with $m_{j}$, meaning heavy ions are less confined than light ones \cite{wineland_1998}. We assume the two ions are in equilibrium, meaning the Coulomb force experienced by each ion combines with the trap to result in net zero force. This mutually repulsive Coulomb potential also contributes to the quadratic potential experienced by each ion (see Appendix \ref{sec:coulomb_expansion}).

\subsection{Phonon rapid adiabatic passage}
\subsubsection*{Two-modes}\label{sec:two_mode_phrap_theory}
We initially restrict our study to two modes $a$ and $b$. If we adjust the trap's DC electrodes slowly relative to the `uncoupled' mode frequencies $\omega_{a(b)}$, the ions will adiabatically follow their equilibrium positions. Similarly, if we ensure $\dot{\omega}_{a(b)}\ll \omega^{2}$ the modes will not squeeze \cite{meekhof_1996,chen_2010}, and we can represent the uncoupled Hamiltonian as:
\begin{eqnarray}\label{eq:uncoupled_two_mode_ladder}
    \hat{H}_{ab} = \hbar\omega_{a}\hat{a}^{\dagger}\hat{a}+\hbar\omega_{b}\hat{b}^{\dagger}\hat{b}.
\end{eqnarray}
Remaining agnostic to its source (see Sec.~\ref{sec:mode_coupling}), if we manipulate the crystal to couple the two modes, the equation becomes:
\begin{eqnarray}\label{eq:uncoupled_two_mode_ladder}
    \hat{H}_{ab}^{\prime} \!=\! \hbar\omega_{a}\hat{a}^{\dagger}\hat{a}\!+\!\hbar\omega_{b}\hat{b}^{\dagger}\hat{b}\! +\!\frac{\hbar\Omega}{2}\Big(\hat{a}^{\dagger}\hat{b}+\hat{a}\hat{b}^{\dagger}\Big).
\end{eqnarray}
We can diagonalize $\hat{H}_{ab}^{\prime}$ with the transformation \cite{agarwal_2012}: \begin{eqnarray}\label{eq:diagonal_two_mode}
    \tilde{H}_{ab}^{\prime} &=& \hat{T}^{\dagger}\hat{H}_{ab}\hat{T}+i\hbar\dot{\hat{T}}^{\dagger}\hat{T},
\end{eqnarray}
where
\begin{eqnarray}\label{eq:rotation_phrap_unitary}
    \hat{T}=\exp\Big(\theta_{r}\Big[\hat{a}^{\dagger}\hat{b}-\hat{a}\hat{b}^{\dagger}\Big]\Big),
\end{eqnarray}
and $\theta_{r}=\arctan(\Omega/2\delta)$, where $\delta\equiv \omega_{a}-\omega_{b}$, giving:
\begin{eqnarray}\label{eq:two_mode_diag_ham}
    \tilde{H}_{ab}^{\prime} &=& \hbar\Big(\omega_{a}\cos^{2}[\theta_{r}]+\omega_{b}\sin^{2}[\theta_{r}] - \frac{\Omega}{2}\sin[2\theta_{r}]\Big)\tilde{a}^{\dagger}\tilde{a} \nonumber \\
    && \hbar\Big(\omega_{b}\cos^{2}[\theta_{r}]+\omega_{a}\sin^{2}[\theta_{r}] + \frac{\Omega}{2}\sin[2\theta_{r}]\Big)\tilde{b}^{\dagger}\tilde{b} \nonumber \\
    &&+i\hbar\dot{\theta}_{r}\Big[\tilde{a}^{\dagger}\tilde{b}-\tilde{a}\tilde{b}^{\dagger}\Big] \nonumber \\
    &\simeq & \hbar\tilde{\omega}_{a}\tilde{a}^{\dagger}\tilde{a}+\hbar\tilde{\omega}_{b}\tilde{b}^{\dagger}\tilde{b}.
\end{eqnarray}
In the second line, we introduced the `coupled' mode frequencies $\tilde{\omega}_{a(b)}$ and assumed $|\dot{\theta}_{r}|\ll \tilde{\delta}$. If we ensure adiabaticity in this way, the time propagator for the system converges to $\hat{T}$\textemdash up to a phase shift that we ignore since it does not affect population transfer. \\

\begin{figure}[b]
\includegraphics[width=0.5\textwidth]{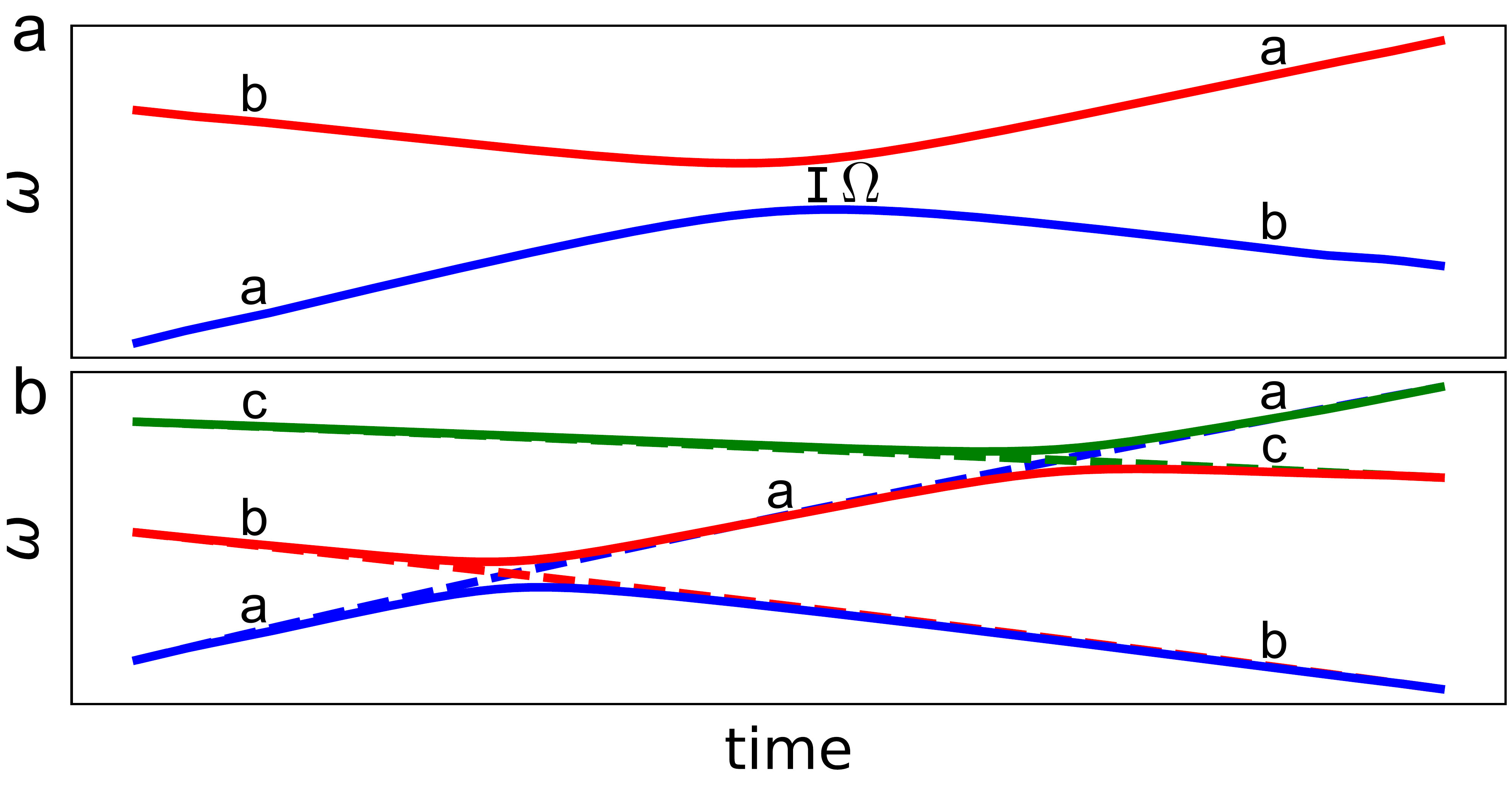}
\centering
\caption{Eigenfrequencies $\omega$ versus time, both in arbitrary units. (a) Example two-mode phrap sequence that exchanges modes $a$ and $b$. The minimum separation of the two frequencies is given by the mode coupling strength $\Omega$ (see text). (b) Example three-mode phrap sequence. By adiabatically ramping through this sequence when mode $a$ is coupled to modes $b$ and $c$ (solid lines), then reversing the process with the coupling off (dashed lines), the operation will permute the mode populations: $a\rightarrow b\rightarrow c\rightarrow a$.}
\label{fig:example_phrap_abc}
\end{figure}

The phrap control sequence begins with $\omega_{a(b)}<\omega_{b(a)}$ and $|\Omega| \ll |\delta|$, such that $\theta_{r}\simeq 0$. During the operation, we increase the size of $\Omega$ to a maximum while ramping through the point $\omega_{a}\simeq \omega_{b}$, i.e. $\delta \simeq 0$. Looking at Eq.~(\ref{eq:two_mode_diag_ham}), we see that when $\delta= 0$, the two modes are split by $\tilde{\delta} = \Omega$\textemdash an avoided crossing. We illustrate this in Fig.~\ref{fig:example_phrap_abc}(a), where we can see the minimum splitting of the two eigenfrequencies is $\sim \Omega$. Finally, we continue to ramp the crystal until $\omega_{a(b)}>\omega_{b(a)}$, while decreasing the mode coupling until $|\Omega| \ll |\delta|$ again. At this point in the sequence, the value of $\theta_{r}\rightarrow \pi/2$ \footnote{It is necessary for the value of $\delta$ to change signs during the operation. If we have ramp $\Omega\rightarrow 0$ at the end of our operation, the value of $\theta_{r}$ has two solutions: $0$ and $\pi/2$. If, when we do this, the value of $\delta$ has not changed signs then the value of $\theta_{r}$ will return to its initial value of $0$ and no transition will occur. On the other hand, if $\delta$ changes signs before $\Omega\rightarrow 0$, the value of $\theta_{r}$ will converge to $\pi/2$.}, and thus the operator $\hat{T}$ exchanges the two modes:
\begin{eqnarray}\label{eq:mode_exchange_two}
    \hat{T}^{\dagger}\hat{a}\hat{T} &\rightarrow& \tilde{b}  \nonumber \\
    \hat{T}^{\dagger}\hat{b}\hat{T} &\rightarrow& \tilde{a}.
\end{eqnarray}
At the same time, Eq.~(\ref{eq:two_mode_diag_ham}) indicates that when $\theta_{r}\rightarrow \pi/2$, $\tilde{\omega}_{a(b)}\rightarrow \omega_{b(a)}$, pointing again to the modes' exchange of character.

\subsubsection*{$N$-modes}\label{sec:n_mode_phrap_theory}

A crystal with $K$ ions will have $N=3K$ motional modes, many of which may be coupled at once. Assuming again that the crystal is adjusted slowly relative to each mode frequency $\omega_{m}$, the `coupled' Hamiltonian becomes:
\begin{eqnarray}\label{eq:multi_mode_coupled_ladder}
    \hat{H}_{N}^{\prime}&=& \hbar\sum_{m}\omega_{m}\hat{a}^{\dagger}_{m}\hat{a}_{m} + \hbar\sum_{m <n}\Omega_{mn}\Big(\hat{a}^{\dagger}_{m}\hat{a}_{n}+ \hat{a}_{m}\hat{a}^{\dagger}_{n} \Big), \nonumber \\
\end{eqnarray}
where $\Omega_{mn}$ is the coupling amplitude between modes $m$ and $n$. Assuming the unitary transformation $T_{N}$ diagonalizes $\hat{H}_{N}^{\prime}$, we can rewrite the equation as:
\begin{eqnarray}\label{eq:phrap_n_modes_diag}
    \tilde{H}_{N}^{\prime} &=& \hat{T}_{N}^{\dagger}\hat{H}_{N}\hat{T}_{N} + i\hbar\dot{\hat{T}}_{N}^{\dagger}\hat{T}_{N} \nonumber \\
    &\simeq & \hbar\sum_{m}\tilde{\omega}_{m}\tilde{a}^{\dagger}_{m}\tilde{a}_{m}, 
\end{eqnarray}
where $\tilde{\omega}_{m}$ are the `coupled' mode frequencies. In the second line, we assume the system changes adiabatically, i.e. that the eigenvectors of the system change slowly compared the mode-mode splittings. Similar to what was shown for two modes, if the frequency splitting $\delta_{mn}\equiv \omega_{m}-\omega_{n}$ of two modes changes sign when $\Omega_{mn}\neq 0$, they exchange character and the values of $\tilde{\omega}_{m}$ and $\tilde{\omega}_{n}$ undergo an avoided crossing. Now, if each value of $\Omega_{mn}$ is adiabatically ramped on/off at the beginning/end of an operation, the system's eigenmodes and eigenfrequencies must converge to what they would be had there been no mode-mode coupling to begin with. Therefore, even if a sequence contains several avoided crossings, it is straightforward to determine the mode permutation induced by a sequence using the eigenfrequency plots with/without mode-mode coupling; when the value of $\tilde{\omega}_{m}$ converges to some $\omega_{n}$ at the end of a sequence, $m$ indicates the initial mode, and $n$ the mode it is mapped onto. See, for example, Fig.~\ref{fig:example_phrap_abc}(b), where we show a three mode (labeled $a$, $b$, and $c$) phrap sequence where $\omega_{a}$ crosses both $\omega_{b}$ and $\omega_{c}$, shown with (solid) and without (dashed) non-zero values of $\Omega_{ab}$ and $\Omega_{ac}$. At the end of the sequence, we see that $\tilde{\omega}_{a}\simeq\omega_{b}$, $\tilde{\omega}_{b}\simeq\omega_{c}$, and $\tilde{\omega}_{c}\simeq\omega_{a}$; this means mode $a$ converges to mode $b$, mode $b$ converges to mode $c$, and mode $c$ converges to mode $a$. The sequence will thus permute mode temperatures according to $\bar{n}_{a}\rightarrow\bar{n}_{b}\rightarrow\bar{n}_{c}\rightarrow \bar{n}_{a}$. Alternatively, we could track the avoided crossings and the mode exchange associated with each.

\subsection{Radial electric fields for mode-mode coupling}\label{sec:mode_coupling}

In this section, we describe how quasi-static radial electric fields applied to a mixed-species crystal generate axial/radial mode-mode coupling. We assume each ion $j$ is confined by $V_{\text{t},j}$, defined in Eqs.~(\ref{eq:paul_trap_unrot}-\ref{eq:potentials}), making the `uncoupled' potential $V_{T}=V_{\text{t},1}+V_{\text{t},2}+V_{\text{C}}$, where:
\begin{eqnarray}\label{eq:coul_hess_diag}
    V_{\text{C}} \simeq  \frac{1}{2}V_{\text{DC}}\Big(x_{d}^{2}-\frac{1}{2}[y_{d}^{2}+z_{d}^{2}]\Big),
\end{eqnarray}
is the Coulomb potential Taylor expanded about the ions' equilibrium positions $\vec{r}_{1(2)}=0$ (see Appendix \ref{sec:coulomb_expansion}), and $\vec{r}_d\equiv \vec{r}_{2}-\vec{r}_{1}$ is their relative displacement in these coordinates. For mode-mode coupling, we ramp on a radial electric field $V_{E}$ in addition to $V_{T}$. This is given by:
\begin{eqnarray}\label{eq:electric_fields}
    V_{E} &=& F_{y}\Big(y_{1}+y_{2}\Big) + F_{z}\Big(z_{1}+z_{2}\Big),
\end{eqnarray}
which we can rewrite as:
\begin{eqnarray}\label{eq:electric_fields}
    V_{E} &=& F_{y,c}q_{y,c} + F_{y,s}q_{y,s} + F_{z,c}q_{z,c} + F_{z,s}q_{z,s},
\end{eqnarray}
where $q_{r,s(c)}$ is the mass-weighted position coordinate of the rSTR(rCOM) eigenmode of $V_{T}$ and:
\begin{eqnarray}\label{eq:force_proj_mode}
    F_{r,s(c)} &\equiv & F_{r}\Big(\frac{\beta_{1}^{r,s(c)}}{\sqrt{m_{1}}} + \frac{\beta_{2}^{r,s(c)}}{\sqrt{m_{2}}}\Big),
\end{eqnarray}
is the force projected onto each mode. Written in this basis, the potential is a set of uncoupled harmonic oscillators, so the changes in the ions' equilibrium positions are given by:
\begin{eqnarray}\label{eq:mode_equil_pos}
    \bar{q}_{r,s(c)} = \frac{F_{r,s(c)}}{\omega_{r,s(c)}^{2}},
\end{eqnarray}
which we can convert to individual ion coordinates $\bar{r}_{1(2)}$, if needed. We can now rewrite $V_{T}$ in these equilibrium positions $r^{~\prime}_{1(2)}=r_{1(2)}-\bar{r}_{1(2)}$. The transformation $V_{T}\rightarrow V_{T}^{\prime}$ eliminates linear terms from the potential, by definition, which means $V_{E}$ vanishes from $V_{T}^{\prime}$; since $V_{\text{t},j}$ is quadratic, $V_{\text{t},j}^{\prime}$ is identical to $V_{\text{t},j}$ except for (also vanishing) linear terms. The Coulomb potential, by contrast, is anharmonic and contributes additional quadratic terms to $V_{T}^{\prime}$. If our applied field displaces the ions in radial coordinate $r$, the expansion of the Coulomb potential gains $\propto x_{d}r_{d}$ terms:
\begin{eqnarray}
    \frac{\partial^{2} V_{\text{C}}}{\partial 
    x_{d}\partial r_{d}}\Big|_{\vec{r}_{1}^{\prime}=\vec{r}_{2}^{\prime}=0} \simeq \frac{3\bar{r}_{d}}{2d}V_{\text{DC}},
\end{eqnarray}
where $d$ is the axial separation of the ions \footnote{If $y_{d}\neq 0$ and $z_{d}\neq 0$, simultaneously, the Coulomb potential will also couple the two radial directions, but these will be small when $y_{d}(z_{d})\ll d$, which we assume.}. Therefore, the total potential defined by $\vec{r}^{~\prime}_{1(2)}=0$ is:
\begin{eqnarray}\label{eq:coulomb_hess}
    V_{T}^{\prime} = V_{T}+V_{\text{C},xr},
\end{eqnarray}
where:
\begin{eqnarray}\label{eq:coulomb_hess_coupling}
    V_{\text{C},xr} \simeq \frac{3V_{\text{DC}}}{2d}[\bar{y}_{d}y_{d} + \bar{z}_{d}z_{d}]x_{d},
\end{eqnarray}
which couples the axial and radial directions of the crystal. Finally, we can rewrite $V_{\text{C},xr}$ in the eigenmode basis by substituting:
\begin{eqnarray}\label{eq:mode_to_coord_sep}
    r_{d}^{\prime} \!\equiv \! \Big(\Big[\frac{\beta^{c}_{2}}{\sqrt{m_{2}}}\! - \!\frac{\beta^{c}_{1}}{\sqrt{m_{1}}}\Big]q_{r,c}\! +\! \Big[\frac{\beta^{s}_{2}}{\sqrt{m_{2}}}\!-\!\frac{\beta^{s}_{1}}{\sqrt{m_{1}}}\Big]q_{r,s}\Big).
\end{eqnarray}
 \\

In same-species crystals, $m_{1}=m_{2}$ and the rSTR modes are antisymmetric ($\beta^{y(z),s}_{1}=-\beta^{y(z),s}_{2}$). Because of this, Eq.~(\ref{eq:force_proj_mode}) implies zero-net force on each rSTR mode, meaning $\bar{q}_{r,s}=0$ according to Eq.~(\ref{eq:mode_equil_pos}). At the same time, same-species rCOM modes are symmetric ($\beta^{y(z),c}_{1}=\beta^{y(z),c}_{2}$), so Eq.~(\ref{eq:mode_to_coord_sep}) indicates $r_{d}^{\prime}=0$, from which it follows that $V_{\text{C},xr}=0$. In other words, radial electric fields cannot generate mode-mode coupling for same-species crystals. In mixed-species crystals, however, $m_{1}\neq m_{2}$ and $\beta^{y(z),c}_{1}\neq \beta^{y(z),c}_{2}$, so applying a uniform force $F_{y(z)}$ to the crystal leads to non-zero $y_{d}(z_{d})$, and, in general, couples axial modes with radial modes.

\subsection{Potential control errors}
\subsubsection{Backphraps}\label{sec:backphraps}

We have discussed the transfer of thermal population from one mode to another, but in a manner that leaves the crystal in a different well than it started in. Assuming our cooling lasers have been configured to the initial well, we must return to the original well without undoing the transfer. If we simply invert the process described above, returning to the initial well slowly compared to all values of $\tilde{\omega}_{m}-\tilde{\omega}_{n}$, then we will undo the transfer; in other words, $\hat{R}\simeq \hat{T}^{\dagger}$ will be the return propagator (up to a phase). To avoid this issue, which we will refer to as a \textit{backphrap}, we must ensure $\hat{R}\rightarrow \hat{I}$ to a sufficient degree. We do this by first ramping off the radial electric field\textemdash giving the mode spectrum shown by the dashed lines in Fig.~\ref{fig:example_phrap_abc}(b) in the absence of noise. This would be sufficient to prevent backphraps under ideal conditions, but stray electric fields will always add non-zero axial/radial coupling. Therefore, we also return to original well over a short timescale relative to initial ramp to reduce the sensitivity to stray fields. In practice, we find typical stray field induced coupling strengths to be small compared any value of $\omega_{m}$, allowing us to return over a timescale short enough to be diabatic with respect to potential backphraps, but slow enough to prevent coherently displacing or squeezing the crystal.

\subsubsection{Parasitic phraps}\label{sec:para_phraps}

The ability to return to the original well over relatively small timescales makes it possible to implement phrap in a way that is insensitive to stray field induced backphraps, preventing heat from leaking into previously-cooled modes. When we use phrap to cool multiple modes at once, however, we must iteratively transfer heat out of targeted modes, without leaking heat into cooled modes\textemdash diminishing overall scheme's efficacy. In any phrap protocol, however, if $\omega_{m}$ and $\omega_{n}$ cross for any two modes, then \textit{any} stray field that introduces a non-zero $\Omega_{mn}$ between the two modes will partially exchange their populations. Since this occurs during the `adiabatic transfer' part of the scheme, the system could be more sensitive to this than backphraps. We refer to the issue as a \textit{parasitic phrap}. \\

While it is not possible to avoid unwanted mode-mode coupling entirely, it is possible to implement phrap schemes in a manner that is insensitive to parasitic phraps. The first way to do this relies on the fact that parasitic phraps are only problematic when they occur between hot modes and cold modes. For example, if modes $a$ and $b$ are partially exchanged, but they are both cold ($\bar{n}_{a}=\bar{n}_{b}\simeq 0$) or they are both hot ($\bar{n}_{a}\sim\bar{n}_{b}\gg 0$), then mixing the two modes will not affect their final temperatures. If, on the other hand, we have cooled mode $a$ to the ground state and left mode $b$ hot, then a parasitic phrap could leak heat into mode $a$; this leakage will increase mode $a$'s final temperature. Therefore, one way to make a scheme insensitive to parasitic phraps is to order the cooling sequence such that stray fields only couple cold/cold or hot/hot modes. Another way to avoid parasitic phraps is to explicitly couple all pairs of modes that cross and incorporate the resulting mode permutation into the overall cooling cycle. We demonstrate multi-mode cooling sequences based on examples of both methods in Sec.~\ref{sec:all-radial-cooling}. 

\section{Experiments}\label{sec:experiments}

\subsection{Setup}\label{sec:experimental_setup}
We conduct all of our experiments with a cryogenic surface-electrode trap, detailed in Ref.~\cite{pino_2021}. Briefly, ${}^{171}\mathrm{Yb}^+$ and ${}^{138}\mathrm{Ba}^+$ ions are loaded 67 $\mu \mathrm{m}$ above the trap surface, forming a $\sim 3.5~\mu \mathrm{m}$ long two-ion crystal oriented in the $x$-direction. The ions are confined radially by a pseudopotential created with a $\sim 42~\text{MHz}$ RF drive and axially by a 36-electrode subset of the trap's 200 DC electrodes. To obtain non-zero laser projections onto each mode, we rotate the static confining potential's principal axes $37.5 \degree$ into the $yz$-plane during Doppler and sideband cooling, and during Raman interrogation. In every experiment, we first Doppler cool the crystal, then initialize the $\mathrm{Yb}^+ $ ion to the state $| {}^2S_{1/2}(F=0,m_{F}=0)\rangle$ using standard techniques \cite{Olmschenk2009}. We then sideband cool a selection of modes to their motional ground states, then send a waveform to the DC electrodes to swap its state with a `hot' (Doppler temperature) mode. Afterward, we perform thermometry using motionally-sensitive Raman pulses on the $\mathrm{Yb}^+$ clock transition $|{}^2S_{1/2}(F=0,m_{F}=0)\rangle \rightarrow | {}^2S_{1/2}(F=1,m_{F}=0)\rangle$, using light with single-photon detuning $\Delta/2\pi = -250$ GHz from $|{}^2P_{1/2}(F=0)\rangle$. We tune the two-photon frequency difference to the clock transition or to one of its motional sidebands, then read out ion's spin state. We repeat each experiment/measurement sequence several hundred times. In the following, we refer to the in-phase (out-of-phase) mode of the $x$ axial direction as XCOM (XSTR), and the $r\in\{y,z\}$ radial direction as RCOM (RSTR).

\begin{figure}[t]
\includegraphics{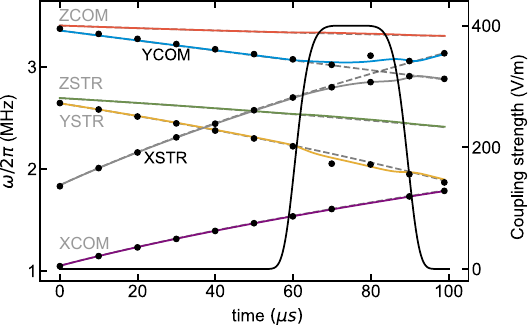}
\caption[Figmodes]{\textbf{Mode coupling demonstration.} Mode frequency profile of a BY crystal undergoing a phrap sequence.  The black dots are measured values, with uncertainties smaller than their marker size. The solid colored lines are a single-parameter fit to our RF voltage.  The crystal is compressed along the axial direction over 100 $\mu $s. Near the YCOM/XSTR degeneracy, we ramp on an electric field along the $y$-direction, making an avoided crossing; the solid black line is the strength of this field versus time. The dashed grey lines represent the mode profiles without the applied field, but with the same axial compression. The black (grey) mode labels indicate the participating (spectator) modes.}
\label{fig:modes}
\end{figure}

\subsection{Mode coupling and transfer demonstrations}\label{sec:mode_coupling_demo}
In Fig. \ref{fig:modes}, we generate axial/radial mode-mode coupling with a radial electric field, as discussed in Sec.~\ref{sec:mode_coupling}. First, we adiabatically compress the crystal along the $x$-direction, bringing XSTR through degeneracy with YCOM. When we near the degeneracy, we smoothly ramp on a $400$ V/m $y$-polarized electric field, pushing the $\mathrm{Ba}^+ $ ($\mathrm{Yb}^+ $) ion 0.7 (1.0) $\mu \mathrm{m}$ off the trap axis. This opens up a $2\pi \times 200\; \mathrm{kHz}$ avoided crossing between the two modes\textemdash large compared to the $1/100\;\mu\mathrm{s}= 10\;\mathrm{kHz} $ inverse waveform time. The solid lines are theoretical predictions, fit only to our RF voltage, and the black dots are Raman spectroscopy measurements; we observe excellent agreement between the two. \\

\begin{figure}[t]
\includegraphics{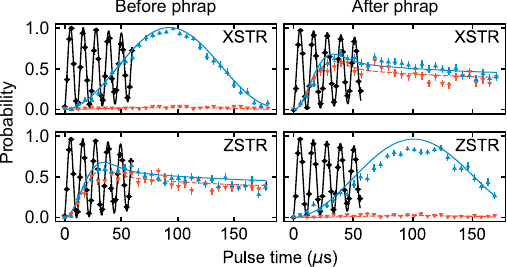}
\caption[Fig2]{\textbf{Transfer demonstration.} Carrier and sideband flops on XSTR and ZSTR, before/after phrap transfer sequence. Pre-phrap, we measure $\bar{n}= 6.5(7) $ on ZSTR and  $\bar{n}= 0.02(1)$ on XSTR. Post-phrap, we measure $\bar{n}= 0.02(1)$ on ZSTR and $\bar{n}= 6.5(8)$ on XSTR. Continuous curves are fits to all data points, assuming thermal population distributions. Black diamonds are pulses on the carrier, and upward pointing blue (downward pointing red) triangles are on the first blue (red) sidebands of the indicated mode. Data is measured and fit simultaneously for XSTR and ZSTR modes. }
\label{fig:demo_zstr}
\end{figure}

In Fig. \ref{fig:demo_zstr}, we demonstrate phrap transfer by exchanging the populations of XSTR and ZSTR. After Doppler cooling, we sideband cool XSTR and YSTR\textemdash cooling the latter prevents parasitic phraps (see Sec.~\ref{sec:para_phraps}). We then couple XSTR with ZSTR using a $400$ V/m $z$-polarized electric field to enable a mode exchange; the induced energy gap is $2\pi\times 300$ kHz. Again, we smoothly ramp on (off) the coupling field before (after) XSTR and ZSTR cross, leaving it off during the return to the initial well; the whole sequence takes $100~\mu\text{s}$. We estimate $\bar{n}$ by flopping on  the carrier and the red/blue sidebands of both modes,  fitting the data to a thermal state. We find that XSTR fully exchanges with ZSTR: fits to our pre-phrap data indicate $\bar{n}= 6.5(7)$ quanta for ZSTR and $\bar{n}= 0.02(1)$ for XSTR, whereas fits to our post-phrap data indicate $\bar{n}=6.5(8)$ quanta for XSTR and $\bar{n}= 0.02(1)$ quanta for ZSTR. \\

\begin{figure}[t]
\includegraphics{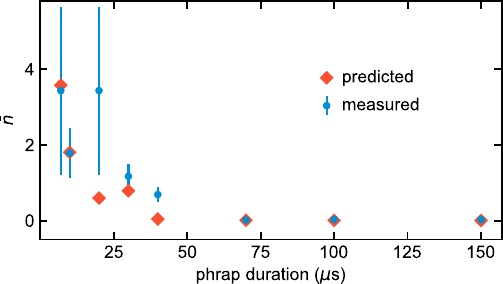}
\caption[Fig3]{\textbf{Residual phonons vs. phrap duration. } Final temperatures $\bar{n}$ of ZSTR as a function of phrap duration using the sequence analyzed in Fig. \ref{fig:demo_zstr}. At shorter times, diabatic behaviour decreases efficiency, causing a corresponding increase of $\bar{n}$. The blue circles are from sideband asymmetry measurements, and the orange diamonds are from numerical simulations. }
\label{fig:speed}
\end{figure}

Next, we determine how fast we can transfer phonons before diabatic behavior decreases the sequence's efficiency. To do this, we vary the time of the phrap sequence analyzed in Fig. \ref{fig:demo_zstr} then run a sideband asymmetry measurement to estimate $\bar{n}$ for ZSTR. We present our results in Fig. \ref{fig:speed}. As expected, we see a decrease in efficiency when the total inverse sequence time gets within an order-of-magnitude of the frequency splitting between XSTR and ZSTR. We observe sub-quanta transfer for all phrap durations over $\sim 50~\mu$s. That the scheme's efficiency converges above a given sequence duration is also indicative of its insensitivity to the size of our applied coupling field; in contrast to techniques based on Rabi flopping \cite{gorman_2014,hou_2022,hou_2023}, so long as the ratio of the induced coupling and the inverse waveform time are kept above a given value, phrap will work. 

\subsection{Robustness to trapping imperfections}\label{sec:exp_robustness}

\begin{figure}[t]
\includegraphics{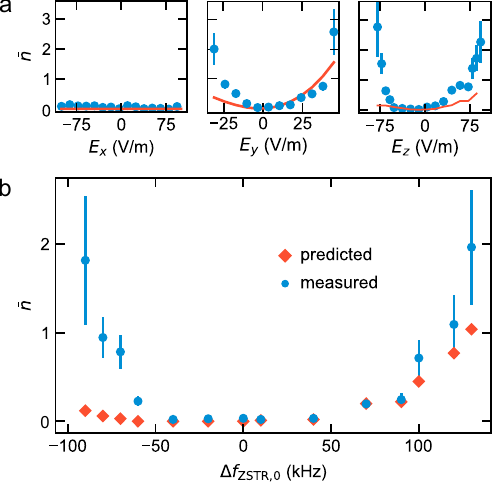}
\caption[Fig4]{\textbf{Phrap tolerances.} Residual temperatures $\bar{n}$ of ZSTR after we apply parameter offsets to the XSTR-ZSTR example transfer sequence. Here, the blue circles are measured values and the orange lines/diamonds are numerical predictions. In part (a), we apply uniform electric fields $E_{r}$ along each direction $r$. In part (b), we alter the initial ZSTR mode frequency by scanning the trap's RF voltage. This frequency range corresponds to changing the RF voltage 13 V.}
\label{fig:tolerances}
\end{figure}

A key advantage of phrap is that its reliance on adiabatic rapid passage, rather than Rabi flopping, makes it insensitive to the precise value of many common experimental parameters. We exemplify this in Figs.~\ref{fig:tolerances}(a) and \ref{fig:tolerances}(b), showing the target mode's residual temperature $\bar{n}$ versus both stray electric fields and mode frequency drifts, respectively. We apply a uniform $z$-polarized coupling field, again using the XSTR-ZSTR phrap sequence analyzed in Figs.~\ref{fig:demo_zstr} and \ref{fig:speed} as an example case; note that, in our prior demonstrations, we sideband cooled YSTR to prevent parasitic phraps from reducing the transfer's efficiency. Here, we allow this possibility by only sideband cooling XCOM and XSTR before each run, leaving the spectator modes at Doppler temperatures \footnote{Throughout these experiments, XCOM remains detuned from the other modes by more than 700 kHz, so it will not cause a parasitic phrap even if its hot.}. \\

In Fig.~\ref{fig:tolerances}(a), we show how a stray $r$-polarized electric field $E_{r}$ can reduce the efficiency of a phrap sequence. We first observe that $\bar{n}$ is independent of $E_{x}$, since its effect is only to translate the trap center along $x$. Examining Fig.~\ref{fig:modes}, however, we can see XSTR crosses YSTR before ZSTR. This means a stray $y$-polarized electric field $E_{y}$ could cause a parasitic phrap, leaking heat into XSTR before we exchange it with ZSTR. In the figure, we see how parasitic phraps between hot and cold modes can add sensitivity to stray electric fields; when $|E_{y}|$ is larger than $\sim 25~$V/m, the parasitic phrap leaks enough heat to prevent sub-quanta cooling. Since XSTR does not cross ZCOM, $E_{z}$ will not also cause parasitic phraps, but it could still decrease efficiency in two ways. Firstly, if $E_{z}$ cancels with the coupling field it will make the system more sensitive to diabaticity. Secondly, as discussed in Sec.~\ref{sec:backphraps}, a large enough $E_{z}$ could cause a backphrap and (partially or fully) reverse the population exchange during the return to the original well. In the figure, we can see the latter is likely the limiting process here, since we begin to see significant residual phonons around $E_{z}\sim \pm 75$~V/m, while the applied coupling field is $400~$V/m.\\

In a second test, we probe phrap's immunity to mode frequency fluctuations by scanning the trap's RF voltage. We design our phrap waveforms assuming a peak of 190 V. Since this value can vary from site to site in QCCD architectures, phrap's insensitivity could help maintain high efficiency when using broadcasted waveforms \cite{malinowski_2023}. In Fig. \ref{fig:tolerances}(b), we vary the RF voltage from 187 V to 200 V, run our example phrap sequence, then perform sideband asymmetry measurements on the ZSTR mode to estimate $\bar{n}$. At either extreme, the phrap begins to fail because the point of degeneracy between XSTR and ZSTR shifts with respect the coupling shim's maximum, making the sequence more sensitive to diabatic behavior. This can be ameliorated, as in Fig. \ref{fig:speed}, by increasing the sequence time to maintain adiabaticity.

\subsection{High-temperature behavior}
\label{sec:breakdown}

\begin{figure}[t]
\includegraphics{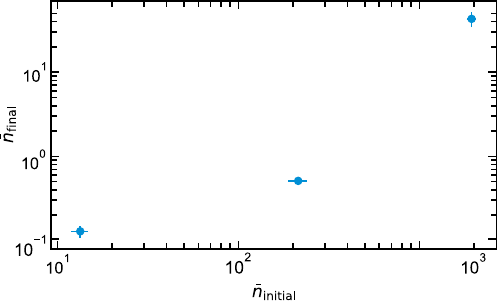}
\caption[Fig4]{\textbf{Residual vs. initial excitation.} Quanta remaining $\bar{n}_{\text{final}}$ in the ZSTR mode after an XSTR-ZSTR phrap sequence, as a function of initial quanta $\bar{n}_{\text{initial}}$. We inject energy into the ZSTR mode via a tone on the RF rails.}
\label{fig:breakdown}
\end{figure}

Although we have focused on transferring excitation out of modes that have been Doppler cooled, phrap could be used to cool hotter modes as well; here we test phrap's temperature limits on an XSTR-ZSTR phrap sequence. In contrast to the experiments described in Figs.~\ref{fig:modes}-\ref{fig:tolerances}, we choose our initial well such that ZSTR is lower frequency than YSTR. This allows us to cross only XSTR and ZSTR, eliminating a potential parasitic phrap from YSTR and letting us single out high-temperature effects. After ground state cooling, we offset $\text{Yb}^{+}-\text{Ba}^{+}$ from the RF null and excite the ZSTR mode by injecting a tone onto the RF rails at the ZSTR frequency. We apply the tone for $75~\mu \text{s}$ with variable amplitude. We then ramp the axial trap frequency past the XSTR-ZSTR crossing while coupling the modes with an electric field in the $z$-direction. Shown in Fig.~\ref{fig:breakdown}, we find that, up to $\sim 200$ quanta, the operation leaves ZSTR at sub-quanta temperatures. For our hottest run, we inject $\sim 2000$ quanta into ZSTR and measure that 97\% of the initial excitation has transferred out of ZSTR. In principle, this means we could use phrap for Doppler cooling\textemdash in the same way we used it for sideband cooling in this work.

\subsection{Cooling all radial modes}\label{sec:all-radial-cooling}

So far, we have shown how to use phrap to cool a mode to near the ground state without directly sideband cooling it. If we can iteratively perform similar transfers on other modes, without leaking heat into modes we have already cooled, it follows that we only need to directly cool a subset of modes to cool an entire crystal. To demonstrate this, we cool all six modes of a $\text{Ba}^{+}\!-\!\text{Yb}^{+}$ crystal while directly sideband cooling only its axial modes. We show this using two distinct cooling sequences, each avoiding parasitic phraps with one of the two methods discussed in Sec.~\ref{sec:para_phraps}. \\

\begin{figure}[t]
\includegraphics{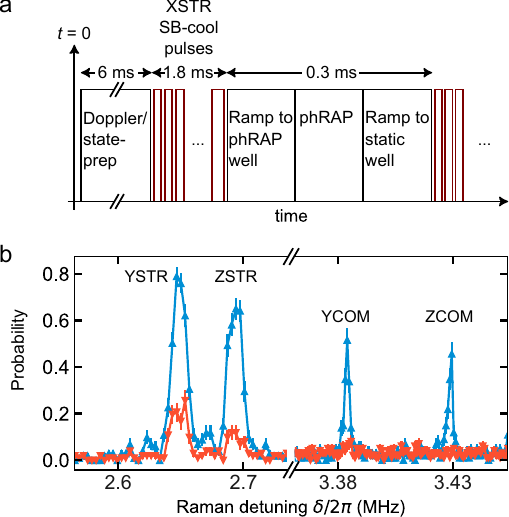}
\caption[Fig6]{\textbf{Cooling all radial modes with phrap.} (a) Timing diagram of a phrap sequence. First, we Doppler cool the $\text{Ba}^{+}-\text{Yb}^{+}$ crystal, initialize $\text{Yb}^{+}$ to $\ket{^{2}S_{1/2}(F=0,m_{F}=0)}$, and sideband cool XCOM and XSTR. To cool a given radial mode, we interpolate from the rotated well to the initial phrap well, implement a phrap exchange between XSTR and a specific radial mode, then interpolate back to the rotated well and cool XSTR (now at Doppler temperature); we do this for all four radial modes. (b) Sideband scan of the targeted modes at the end of the sequence. We invert the $x$-axis of the red sideband data (red triangles pointing down) to overlap with that of the blue sideband data (blue triangles pointing up), adding lines to guide the eye. According to the sideband ratios, all modes are sub-quanta.}
\label{fig:allradials}
\end{figure}

In the first demonstration, we cool the crystal with four distinct phrap sequences. As with previous experiments, we begin by Doppler cooling every mode and initializing $\mathrm{Yb}^+$. Next, we sideband cool XSTR to $<\!0.1$ quanta, interpolate from the rotated static well to the initial phrap well in 100 $\mu\text{s}$, run the phrap sequence in 100 $\mu\text{s}$ to exchange XSTR with one of the radial modes, and finally interpolate back to the rotated well in 100 $\mu\text{s}$. We repeat this sequence for each radial mode. We provide a timing diagram for this in Fig. \ref{fig:allradials}(a). \\

 We phrap the radial modes from smallest to largest frequency, i.e. (YSTR, ZSTR, YCOM, ZCOM), to minimize the impact of parasitic phraps. At the end of the sequence, we probe the red and blue sidebands of each radial mode\textemdash shown in Fig. \ref{fig:allradials}(b). From the red/blue sideband ratios, we estimate the final mode temperatures to be $(\bar{n}_{\mathrm{YSTR}},\bar{n}_{\mathrm{ZSTR}},\bar{n}_{\mathrm{YCOM}},\bar{n}_{\mathrm{ZCOM}}) = (0.5(1), 0.22(8), 0.08(5),0.05(3))$. The residual excitation in the radial STR modes at the end of the sequence is likely due to photon recoil during the XSTR cooling step, not from the phrap transfer itself (see Appendix \ref{app:prep_heating}). The degree to which we ground state cool the radial COM modes is of particular note: directly cooling these modes would be onerous, as they are decoupled from the coolant ion \cite{wubbena_2012,sosnova_2021}. Traditional sideband cooling would require 10s of milliseconds and sub-kHz mode stabilities. \\

\begin{figure}[t]
\includegraphics{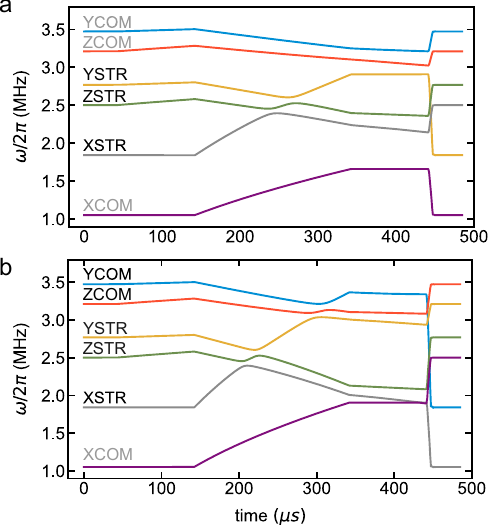}
\caption[Fig6]{\textbf{Permutation phraps.}  Mode frequencies versus time for the permutation phrap sequences we use to cool the radial modes of a $\text{Ba}^{+}\!-\!\text{Yb}^{+}$ crystal. At the end of either sequence, XSTR and every mode it crosses (black labels) cyclically permute, while the other modes (grey labels) act as spectators; after each permutation, we sideband cool XSTR. Implementing (a) twice cools the radial STR modes, after which implementing (b) twice cools the radial COM modes. We measure $<0.5$ quanta in all four radial modes at the end of the sequence. }
\label{fig:staticphrap}
\end{figure}

In the above sequence, we cooled the radial modes by iteratively exchanging XSTR with a given radial mode, each exchange requiring a distinct waveform. Alternatively, we could leverage a phrap sequence that intentionally couples several modes at once, setting up a multi-mode permutation (described in Sec.~\ref{sec:n_mode_phrap_theory}). As discussed in Sec.~\ref{sec:para_phraps}, such a sequence should be insensitive to stray fields as long as they are smaller than the applied coupling fields. For this sequence, we adiabatically turn on $200~$V/m coupling fields in both radial directions for $140\;\mu$s before the axial ramping step, coupling XSTR to every radial mode. We then axially compress the crystal to two extents: Fig.~\ref{fig:staticphrap}(a), where XSTR crosses only the radial STR modes; and Fig.~\ref{fig:staticphrap}(b), where XSTR crosses every radial mode. After either operation, XSTR and every mode it crosses will cyclically permute, i.e. $\bar{n}_{\text{XSTR}}\rightarrow\bar{n}_{\text{YSTR}}\rightarrow\bar{n}_{\text{ZSTR}}\rightarrow\bar{n}_{\text{XSTR}}$ for Fig.~\ref{fig:staticphrap}(a) and $\bar{n}_{\text{XSTR}}\rightarrow\bar{n}_{\text{YSTR}}\rightarrow\bar{n}_{\text{ZSTR}}\rightarrow\bar{n}_{\text{YCOM}}\rightarrow \bar{n}_{\text{ZCOM}}\rightarrow\bar{n}_{\text{XSTR}}$ for  Fig.~\ref{fig:staticphrap}(b). We ground-state cool XSTR after each permutation. In short, applying the Fig.~\ref{fig:staticphrap}(a) phrap twice will cool the radial STR modes, and then applying the Fig.~\ref{fig:staticphrap}(b) phrap twice will cool the radial COM modes. After sequentially applying all four phraps, we again use the sideband ratios to estimate the final radial mode temperatures, finding $(\bar{n}_{\mathrm{YSTR}},\bar{n}_{\mathrm{ZSTR}},\bar{n}_{\mathrm{YCOM}},\bar{n}_{\mathrm{ZCOM}}) = (0.19(4), 0.38(7), 0.3(1),0.5(2))$. Similar to the previous sequence, we attribute the residual temperatures primarily to photon recoil heating the radial STR modes during the XSTR cooling step (see Appendix~\ref{app:prep_heating}); unlike the previous sequence, the heat permutes to the radial COM modes.

\section{Generalized all mode cooling}\label{sec:gen_all_mode_cooling}

\begin{figure}[b]
\includegraphics[width=0.5\textwidth]{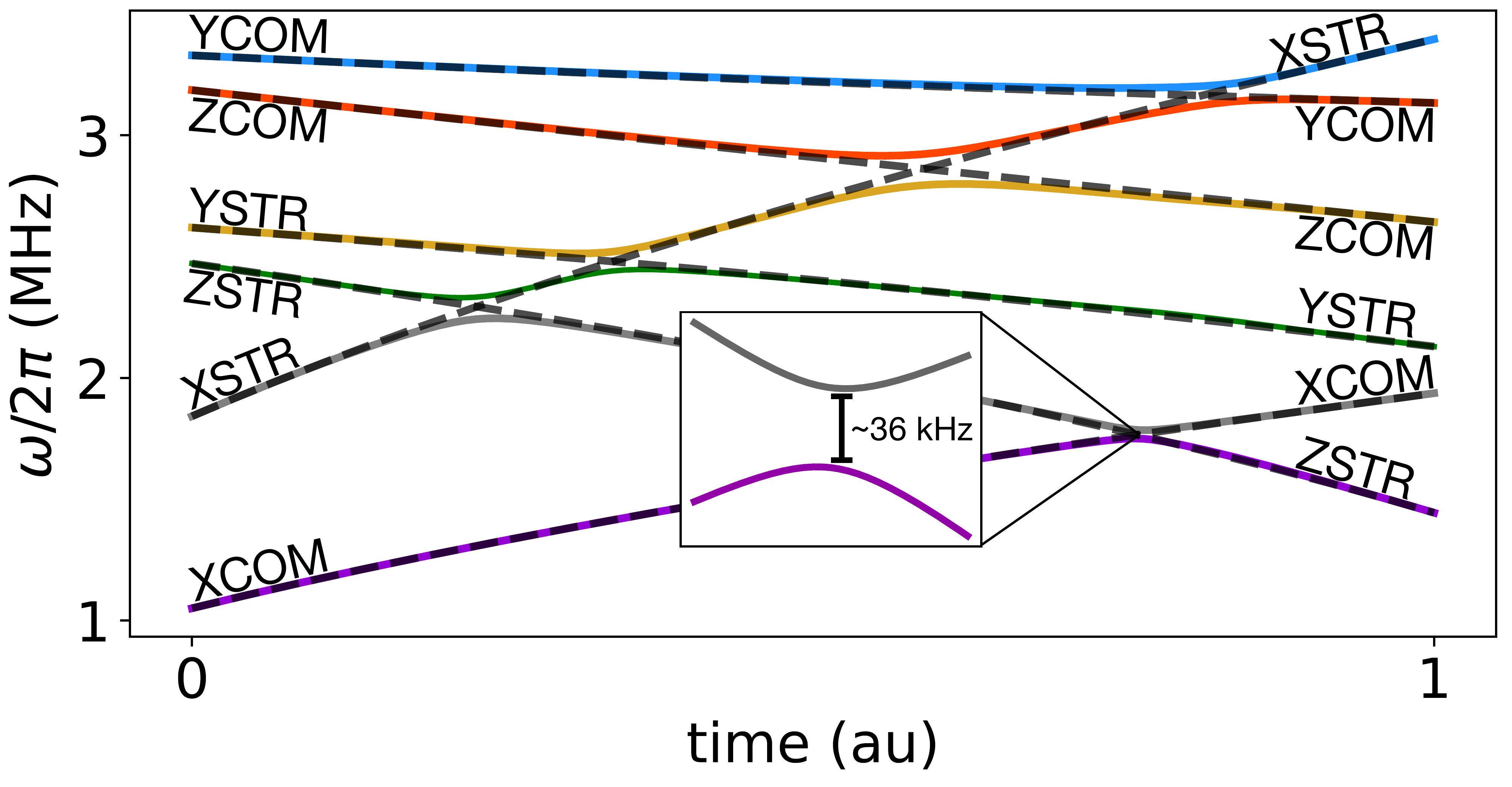}
\centering
\caption{Mode frequencies versus time (arbitrary units) for a phrap sequence that permutes \textit{every} mode of a $\text{Ba}^{+}-\text{Yb}^{+}$ crystal. We couple the modes by sinusoidally ramping on radial electric fields to $E_{y}=E_{z}=200$ V/m for the first and last 1/10 of the sequence. The effect of is to permute every mode in the crystal, mapping the motional populations $\bar{n}_{\text{XCOM}}\rightarrow\bar{n}_{\text{ZSTR}}\rightarrow\bar{n}_{\text{YSTR}}\rightarrow\bar{n}_{\text{ZCOM}}\rightarrow\bar{n}_{\text{YCOM}}\rightarrow\bar{n}_{\text{XSTR}}\rightarrow\bar{n}_{\text{XCOM}}$. Given the ability to directly cool a single crystal mode, this sequence enables all-mode-cooling. The inset shows the smallest $\delta/2\pi\simeq 36~\text{kHz}$ avoided crossing, setting the condition for adiabaticity for the sequence.}
\label{fig:radial_e_field}
\end{figure}

As we have now shown, phraps between multiple modes will cyclically permute a subset of a trapped ion crystal's modes. Now, lets assume a cooling sequence has two steps: directly cooling one mode of an $n$-mode subset, then permuting the subset with phrap. After $n$ repetitions, every mode in the subset will be cold. It follows that if a phrap cyclically permutes all $N$ modes of a crystal, then directly cooling \textit{any} mode will cool the entire crystal upon $N$ repetitions. In Fig.~\ref{fig:radial_e_field}, we show the (theoretical) mode frequency plot for such a sequence. In this example, we assume a $\text{Yb}^{+}-\text{Ba}^{+}$ crystal radially confined with a value corresponding to a single-$^{171}\text{Yb}^{+}$ radial frequency of $\omega_{r}/2\pi=2.75~\text{MHz}$. Initially, we confine the ions axially with a DC potential $V_{DC}=V_{DC,0}\Big(x^{2}-[y^{2}+z^{2}]/2 \Big)$, where $V_{DC,0}=m_{\text{Yb}}\omega_{x}^{2}/2$ and $\omega_{x}/2\pi=1~\text{MHz}$. Additionally, we split the radial confinement using a $yy-zz$ field, so that the $z$-direction modes are lower frequency than the $y$-direction modes. During the operation, we increase the axial confinement while ramping on radial electric fields $E_{y}=E_{z}=200~\text{V/m}$. This causes the axial modes to increase in frequency and the radial modes to decrease in frequency to the extent the XSTR mode crosses every radial mode, and the XCOM mode crosses the lowest frequency radial mode YSTR. If we scan through each avoided crossing slowly enough to maintain adiabaticity, we can apply the technique outlined in Sec.~\ref{sec:n_mode_phrap_theory} to determine the effect of the sequence, which is to map the motional states $\bar{n}_{\text{XCOM}}\rightarrow\bar{n}_{\text{ZSTR}}\rightarrow\bar{n}_{\text{YSTR}}\rightarrow\bar{n}_{\text{ZCOM}}\rightarrow\bar{n}_{\text{YCOM}}\rightarrow\bar{n}_{\text{XSTR}}\rightarrow\bar{n}_{\text{XCOM}}$. Since this phrap sequence permutes every mode in the crystal, combining it with direct cooling on any mode should enable all-mode-cooling. In Fig.~\ref{fig:radial_e_field}, we can see the narrowest avoided crossing is the $\delta/2\pi \sim 35~\text{kHz}$ crossing between YSTR and XCOM, which could be a speed bottleneck. If this is the case, it is possible to increase the avoided crossing's size by adding a $E_{xy}$ rotation shim to the system. We find that doing this can significantly increase the size of the avoided crossing, without significantly decreasing the size of the other crossings. This allows us to operate on much shorter timescales, at cost of introducing ion order dependence to the coupling strength (see Appendix \ref{sec:trap_rotations}).

\section{Conclusion}\label{sec:conclusion}

In this work, we proposed and demonstrated a technique to indirectly cool the motional modes of trapped ion crystals. Due to its simplicity and immunity to common experimental drifts, we anticipate that phrap will play an important role in reducing the time trapped ion quantum computers spend cooling. Similarly, it could simplify cooling in trapped ion atomic clocks. Phrap would be especially impactful when the sideband cooling of certain modes is hard due to beam geometries, or because the modes are decoupled from the coolant ion. The technique has minimal requirements, and should be possible in most trapped ion laboratories that use mixed-species crystals, can measure sidebands, and can scan their trap's DC electrodes. To exchange two modes, simply scan the trap electrodes to find a point where the avoided crossing between the modes is large, and another where it is small; use the former to adiabatically ramp through the crossing and the latter to diabatically return to the original well. Finally, we note that, though our demonstrations focused on two ion crystals, phrap can be applied to $N$ ion crystals as well. 

\section*{Acknowledgements}

We would like to thank M. Foss-Feig, J. Gaebler, B. J. Bjork, C. Langer, M. Schecter, J. Bartolotta, C. N. Gilbreth, and P. J. Lee for helpful discussions.

\appendix 

\section{Heating of motional modes via state preparation}\label{app:prep_heating}
\begin{figure}[h]
\includegraphics[width=0.5\textwidth]{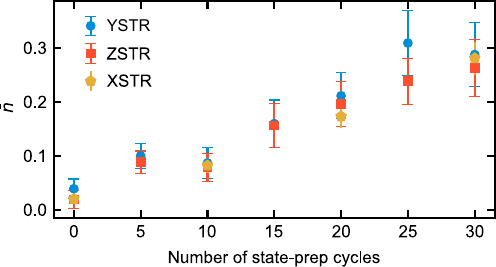}
\centering
\caption{Temperature of motional modes due to photon recoil versus number of state-preparation cycles. In general, our sideband cooling sequence needs one cycle to remove one quanta.}
\label{fig:stateprep}
\end{figure}

For the data presented in Section \ref{sec:all-radial-cooling} of the main text, each phrap operation was followed by sideband cooling on XSTR. We find that photon recoil from our sideband cooling\textemdash specifically, the state-preparation pulse after the red sideband pulse\textemdash can induce a small amount of motional excitation on certain spectator modes. In Fig. \ref{fig:stateprep}, we measure this effect in $\text{Ba}^{+}-\text{Yb}^{+}$ crystals by repeating the state-prep cycle used in the main text $N$ times, then measuring $\bar{n}$ for various modes via fitting to carrier and red/blue sideband flops. Our typical Doppler temperatures, $10-15$ quanta, require approximately as many red-sideband/state-prep cycles on XSTR as there are phonons, implying $\approx 0.1\!-\!0.2$ quanta of excitation will be added to the radial STR modes each time we cool XSTR. If we have already ground state cooled these modes, this will raise their final temperatures.

\section{Paul traps}
Laplace's equation tells us it is not possible trap a charged particle using electrostatic fields, but laws tend to have loopholes. In Paul traps \cite{paul_1990}, we create the positive potential curvature needed for a trap by modulating a harmonic term at a frequency $\Omega_{\text{RF}}$: 
\begin{equation}
    V_{\text{RF}}(t)\simeq V_{\text{RF}}\cos(\Omega_{\text{RF}}t)(y^{2}-z^{2}),
\end{equation}
which gives an effective `pseudopotential' \cite{wineland_1998} for a mass $m_{j}$ ion:
\begin{equation}\label{eq:pseudo}
    V_{\text{pp},j} = \frac{V_{\text{RF}}^{2}}{\Omega_{\text{RF}}^{2}m_{j}}(y^{2}+z^{2}),
\end{equation}
when averaged over time. If the ion chain is relatively cold and near the RF null, $V_{\text{pp},j}$ is often a good approximation for the total confinement, which we assume in the main text. Importantly for this work, $V_{\text{pp},j}$ is inversely proportional to $m_{j}$. While $V_{\text{pp}}$ confines the ions radially, we confine them axially with a DC potential: 
\begin{equation}\label{eq:axial_dc_electrodes}
    V_{\text{DC}} = V_{\text{DC}}(x^{2}-[c_{y}y^{2}+c_{z}z^{2}]),
\end{equation}
where $c_{y} + c_{z} = 1$ to ensure Laplace's equation is satisfied. The total confinement provided by the trap to ion $j$ in each direction is:
\begin{eqnarray}\label{eq:potentials_app}
    V_{xx,j} &=& V_{\text{DC}}, V_{rr,j} = V_{\text{pp},j} - c_{r}V_{\text{DC}},
\end{eqnarray}
for $r\in\{y,z\}$, giving a total potential: 
\begin{eqnarray}\label{eq:paul_trap_unrot_app}
    V_{\text{t},j} = V_{xx}x_{j}^{2} + V_{yy,j}y_{j}^{2} + V_{zz,j}z_{j}^{2},
\end{eqnarray}
for each ion $j$. If we do not use DC electrodes to rotate the crystal or apply transverse E-fields (see main text), and there no trap anharmonicities (assumed for simplicity), Eq.~(\ref{eq:paul_trap_unrot_app}) is a complete description of a trap. While Eq.~(\ref{eq:paul_trap_unrot_app}) implies different radial confinements for different massed ions, for one ion or same-species crystals, the 3D-confinement is identical for each ion\textemdash reducing the number of independent curvatures that describe the system to three; this has implications about the rotational symmetry of same-species traps that are important for this work (see Appendix~\ref{app:rotated_wells}).

\section{Coulomb potential}\label{sec:coulomb_expansion}
For two-ion crystals, both ions separate themselves along the direction of weakest confinement, which we here take to be $x$- (axial) direction. Compared to a single trapped ion, the total potential of the crystal includes the Coulomb interaction:
\begin{equation}\label{eq:coulomb_orig}
    V_{\text{C}}=\frac{kq^{2}}{|\vec{r}_{2}-\vec{r}_{1}|},
\end{equation}
where $k$ is Coulomb's constant, $q$ is the ion charge, and the $1(2)$ subscripts indicate the qubit(coolant) ion. The total potential of a multi-ion system is then $V_{\text{t}} = V_{\text{t},1}+V_{\text{t},2}+V_{\text{C}}$, i.e. the effective potential of the trap plus the ions' mutual Coulomb repulsion. The set of ion equilibrium positions corresponds to the point where each ion experiences net $0$ force:
\begin{eqnarray}\label{eq:equil_condition}
    \frac{\partial V_{\text{t}}}{\partial x_{\alpha}} = 0,
\end{eqnarray}
for each ion $m$. Regardless of their masses, two ions confined in a trap defined by Eq.~(\ref{eq:paul_trap_unrot_app}) will have equilibrium positions centered on the trap axis $y_{1(2)}=z_{1(2)}=0$, with each ion a distance $d/2\equiv\pm (kq^{2}/4V_{\text{DC}})^{1/3}$ from the origin. We define our coordinate system for each ion $\{\vec{r}_{1},\vec{r}_{2}\}$ by each ion's displacement from these positions. \\

\noindent We want to understand the behavior of such a crystal in the presence of an added E-field or a rotation shim control field, each of which, for simplicity, we assume displaces the ions a small distance relative to $d$, i.e. $|\vec{r}_{1(2)}|\ll|\tilde{r}_{1(2)}|$. If we adiabatically ramp-on a control field that brings each ion to a new set of equilibrium positions $\vec{r}_{1(2)}$, i.e. condition Eq.~(\ref{eq:equil_condition}) is satisfied, we can then approximate Eq.~(\ref{eq:coulomb_orig}) with the $2^{\text{nd}}$-order terms in its Taylor expansion. Rewriting Eq.~(\ref{eq:coulomb_orig}) in terms of the ion-ion displacement vector $\vec{r}_{d}= \vec{r}_{2}-\vec{r}_{1}$ and expanding $V_{\text{C}}$ around their equilibrium positions gives: 
\begin{eqnarray}
    \frac{\partial^{2} V_{\text{C}}}{\partial x^{2}_{d}} = V_{\text{DC}}, \frac{\partial^{2} V_{\text{C}}}{\partial y^{2}_{d}} = -\frac{1}{2}V_{\text{DC}}, \frac{\partial^{2} V_{\text{C}}}{\partial z^{2}_{d}} = -\frac{1}{2}V_{\text{DC}}. \nonumber \\
\end{eqnarray}
If the ions remain on the axis of the trap ($y_{d}=z_{d}=0$), this represents the Coulomb potential's contribution to the confinement experienced by each ion:
\begin{eqnarray}\label{eq:coul_hess_diag}
    V_{C,0} \simeq  \frac{1}{2}V_{\text{DC}}\Big(x_{d}^{2}-\frac{1}{2}[y_{d}^{2}+z_{d}^{2}]\Big),
\end{eqnarray}
after which we can substitute $\{\vec{r}_{1},\vec{r}_{2}\}$ when convenient. This term adds curvature to the axial direction, since the repulsive force decreases with $x_{d}$, and subtracts curvature from the radial directions because the repulsive force's radial projection increases with $y_{d}$ and $z_{d}$. Importantly, Eq.~(\ref{eq:coul_hess_diag}) only depends on the displacement of each ion, i.e. $\vec{r}_{d}$. As a result, it tends to affect the frequencies of out-of-phase modes (`stretch' STR modes) more than in-phase modes (`center-of-mass' COM modes). This is why axial STR modes tends to be higher frequency than axial COM modes, while radial STR modes tend to be lower frequency than radial COM modes \cite{wineland_1998}. \\

\section{Motional modes}

When the ion crystal is on the trap axis the total potential of the two ion crystal is:
\begin{eqnarray}\label{eq:v0_individual}
    V_{0} &=& V_{\text{t},1}+V_{\text{t},2}+V_{\text{C},0} \nonumber \\ 
    &=& \sum_{\alpha = 1,2}\Big(V_{xx,\alpha}x_{\alpha}^{2}+ V_{yy,\alpha}y_{\alpha}^{2} + V_{zz,\alpha}z_{\alpha}^{2}\Big) \nonumber \\
    && + \frac{1}{2}V_{\text{DC}}\Big([x_{{2}}-x_{1}]^{2}-\frac{1}{2}[y_{{2}}-y_{1}]^{2}+\frac{1}{2}[z_{{2}}-z_{1}]^{2}\Big) \nonumber \\
    &=& \sum_{r,\alpha,\alpha^{\prime}}V_{rr,\alpha\alpha^{\prime}}r_{\alpha}r_{\alpha^{\prime}},
\end{eqnarray}
where, in the second line, we have grouped the coefficient of each harmonic term and introduced the two-ion Hessian matrix $V_{rr,\alpha\alpha^{\prime}}$. Importantly, this equation represents an \textit{uncoupled} trap, i.e. $V_{rr^{\prime},\alpha\alpha^{\prime}}\propto \delta^{rr^{\prime}}$; later, and in the main text, we will discuss methods for manipulating the crystal to add $V_{rr^{\prime},\alpha\alpha^{\prime}}$ terms, where $r\neq r^{\prime}$\textemdash crucial to this work. This makes the Hamiltonian:
\begin{eqnarray}\label{eq:h0_not_mass_weighted}
    H_{0}&=& \sum_{r,\alpha}\frac{p_{r,\alpha}^{2}}{2m_{\alpha}} + \sum_{r,\alpha,\alpha^{\prime}}V_{rr,\alpha\alpha^{\prime}}r_{\alpha}r_{\alpha^{\prime}}.
\end{eqnarray}
In this coordinate system, there is no orthogonal transformation $T$ that allows us to simultaneously diagonalize the kinetic and potential energy components of $\hat{H}_{0}$. We can, however, rewrite the problem in mass weighted coordinates $\tilde{r}_{\alpha}=\sqrt{m}_{\alpha}r_{\alpha}$:
\begin{eqnarray}\label{eq:h0_mass_weighted}
    \tilde{H}_{0}&=& \frac{1}{2}\sum_{r,\alpha}\tilde{p}_{r,\alpha}^{2} + \sum_{r,\alpha,\alpha^{\prime}}\tilde{V}_{rr,\alpha\alpha^{\prime}}\tilde{r}_{\alpha}\tilde{r}_{\alpha^{\prime}},
\end{eqnarray}
recasting the information about ion mass in $\tilde{V}_{rr,\alpha\alpha^{\prime}} =\sqrt{m_{\alpha}m_{\alpha^{\prime}}} V_{rr,\alpha\alpha^{\prime}}$. In this coordinate system, the kinetic energy term takes the form on an inner product, and, therefore, is invariant under orthogonal transformations; we can now uncouple $\tilde{H}_{0}$ by diagonalizing $\tilde{V}_{rr,\alpha\alpha^{\prime}}$ with an orthogonal matrix $T$. We can now rewrite Eq.~(\ref{eq:h0_mass_weighted}) as system of uncoupled motional modes:
\begin{eqnarray}\label{eq:h0_mass_weighted}
    \tilde{H}_{0,q}&=& \frac{1}{2}\sum_{n}\dot{\tilde{q}}^{2}+\omega^{2}_{n}\tilde{q}^{2}_{n},
\end{eqnarray}
where we have introduced mode position and momentum coordinates $q_{n}$ and $\dot{q}_{n}$, respectively. We can transform between the two coordinate systems using:
\begin{eqnarray}
    r_{\alpha} = \frac{1}{\sqrt{m_{\alpha}}}\sum_{m}\beta^{n}_{r,\alpha}q_{n}
\end{eqnarray}
According to Ehrenfest's theorem, the expectation values of harmonic Hamiltonians, such as $\tilde{H}_{0,q}$, always follow Newton's equations of motion. Since we only consider harmonic interactions here, we can switch between the quantum and classical pictures with impunity, whenever convenient. Specifically, we use classical numerical simulations of the full ion crystal in Sec.~\ref{sec:experiments} because they scale in a more straightforwardly favorable way with system size. While, in Sec.~\ref{sec:theory}, we describe the theory of phrap using ladder operators:
\begin{eqnarray}
    \tilde{q}_{n} &=& \sqrt{\frac{\hbar}{2\omega_{n}}}\Big(\hat{a}^{\dagger}_{n}+\hat{a}_{n} \Big) \nonumber \\
    \dot{\tilde{q}}_{n} &=& i\sqrt{\frac{\hbar\omega_{n}}{2}}\Big(\hat{a}^{\dagger}_{n}-\hat{a}_{n} \Big),
\end{eqnarray}
because it allows us to better isolate the behavior of specific modes using the rotating wave approximation. We can insert these into Eq.~(\ref{eq:h0_mass_weighted}) to give:
\begin{eqnarray}\label{eq:ladders_uncoupled_all}
    \hat{H}_{0,q} = \hbar\sum_{n}\omega_{n}\hat{a}^{\dagger}_{n}\hat{a}_{n}.
\end{eqnarray}\\

\section{Sympathetic cooling}
The modal structure of mixed-species crystals has been discussed in detail \cite{wubbena_2012}, but its relevance to this work is worth pointing out. Each direction now has two modes associated with it: one mode where the ions move in same direction `COM' (center-of-mass-like), and one where the ions mode in different directions (stretch-like) `STR'. Since the ions are separated axially, the Coulomb curvature results in XSTR being higher-frequency than XCOM (see Sec.~\ref{sec:coulomb_expansion}). Since the Coulomb force is a static potential, it must decrease curvature in the radial directions to satisfy Laplace's equation. As a result YSTR and ZSTR are, in general, lower frequency than YCOM and ZCOM, respectively; note that this does not necessarily mean that YSTR is always lower frequency than ZCOM. As discussed in Ref.~\cite{wubbena_2012}, when the value of $m_{1}/m_{2}\not\approx 1$, the radial mode `decouple', meaning they tend to be comprised almost entirely of the coolant or the qubit ion. When $m_{1}>m_{2}$, radial STR modes are slow-cooling and radial COM modes are fast-cooling. When $m_{2}>m_{1}$, the case demonstrated below, radial STR modes are fast-cooling and radial COM modes are slow-cooling.

\section{Axial/radial rotation shims}\label{sec:trap_rotations}

We can directly couple radial and axial modes with a rotation shim:
\begin{eqnarray}\label{eq:axial_radial_rot}
    V_{R} = V_{xy}xy + V_{xz}xz,
\end{eqnarray}
defined about the center of the trap. Because of their axial separation, this will induce a differential force in $y$ or $z$\textemdash the coolant ion experiencing $F_{2}=-V_{xr}d/2$ and the qubit ion experiencing $F_{1}=V_{xr}d/2$ in radial direction $r$, causing the crystal axis to rotate about its center. If $V_{x}$ is positive/negative, this will result in a negative/positive value of $\tilde{r}_{d}$ at the ions' new equilibrium positions, making a $V_{\text{C},xr}$ that cancels with the $V_{\text{rot}}$ that created it. For same-species crystals this canceling is exact (see next Appendix~\ref{app:rotated_wells}). For the mixed-species crystals we explore in the main text, we find that, while the cancellation is not exact, it is difficult to induce significant coupling between axial and radial modes because of it. As we discuss in Sec.~\ref{sec:gen_all_mode_cooling}, we can combine axial/radial rotation shims with the radial electric fields (discussed in the previous subsection) to amplify mode-mode couplings in certain situations. Additionally, in Appendix~\ref{sec:radial_rotations}, we discuss using radial/radial rotation shims, i.e. potential terms $\propto yz$, to exchange the motional state of every radial mode in one direction with its coorespondent mode in the other radial direction.

\section{Rotated wells}\label{app:rotated_wells}

If we apply a rotation shim that couples $x_{j}$ and $x_{k}$, the potential for these two degrees of freedom is:
\begin{eqnarray}\label{eq:unrotated_1_ion_hess}
    V_{\text{rot}} = V_{jj}x_{j}^{2} + V_{kk}x_{k}^{2} + V_{jk}x_{j}x_{k}.
\end{eqnarray}
If we rotate this coordinate system about the direction orthogonal to both directions, i.e. the cross-product of the two unit vectors $\hat{x}_{j}\times\hat{x}_{k}$, by an angle $\theta$ we get:
\begin{eqnarray}\label{eq:rotated_1_ion_hess}
    x_{j} &=& x_{j}^{\prime}\cos(\theta) + x_{k}^{\prime}\sin(\theta)  \\
    x_{k} &=& x_{k}^{\prime}\cos(\theta) - x_{j}^{\prime}\sin(\theta). \nonumber
\end{eqnarray}
Inserting this into Eq.~(\ref{eq:unrotated_1_ion_hess}), we obtain the potential in a general, rotated coordinate system:
\begin{eqnarray}
    V^{\prime}_{\text{rot}}&=& \Big(V_{jj}\cos^{2}[\theta] + V_{kk}\sin^{2}[\theta]  - \frac{1}{2}V_{jk}\sin[2\theta]\Big)x_{j}^{\prime 2} + \nonumber \\
    && \Big(V_{kk}\cos^{2}[\theta] + V_{jj}\sin^{2}[\theta] + \frac{1}{2}V_{jk}\sin[2\theta]\Big)x_{k}^{\prime 2} + \nonumber \\
    && \Big([V_{jj}-V_{kk}]\sin[2\theta] + V_{jk}\cos[2\theta] \Big)x_{j}^{\prime}x_{k}^{\prime}.
\end{eqnarray}
If we choose:
\begin{eqnarray}
    \theta = \frac{1}{2}\arctan\Big(\frac{V_{jk}}{V_{kk}-V_{jj}} \Big),
\end{eqnarray}
the $\propto x_{j}^{\prime}x_{k}^{\prime}$ term vanishes, giving an uncoupled 3D oscillator. In other words, the added $\propto x_{j}x_{k}$ torque term can be interpreted as a rotation of the trapping potential. For same-species crystals, the value of $\theta$ for both ions is identical; meaning, if adiabaticity is maintained, the mode structure of the crystal will not change. In other words, the types of phrap sequences discussed in the main text would not work. This means any phrap protocol used to transfer motional states of same-species crystals must do so globally, uniformly rotating all of the motion in one direction into another (orthogonal) one. Importantly, neither of these symmetry arguments apply to mixed-species crystals because of the $\propto 1/m$ dependence of the pseudopotential; as a result, an E-field will translate each ion different distances, and a harmonic torque term about the trap center by different angles. As we will discuss, this enables many phrap transitions unique to mixed-species crystals.

\section{Simultaneous application of radial electric fields and rotation shims}

\begin{figure}[b]
\includegraphics[width=0.5\textwidth]{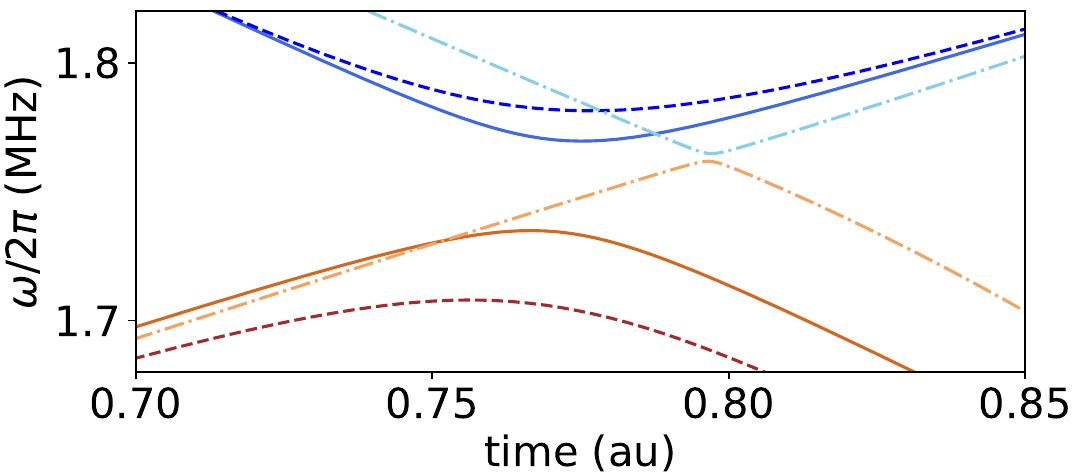}
\centering
\caption{Eigenfrequencies $\omega$ versus time (arbitrary units) for example phrap sequences where we show how rotation shims can combine with radial electric fields to help strengthen mode-mode coupling, at expense of being sensitive to ion ordering, for the YSTR/XCOM crossing shown in the text and inset of Fig.~\ref{fig:radial_e_field}. The solid lines represent the (order-insensitive) case when we couple the two modes with only a radial electric field $|E_y|=200$ V/m. In addition, we add a $\propto xy$ rotation shim $V_{r,0}\simeq 1.75\times 10^{7}$ V/$m^{2}$ to the crystal, which increases the splitting to $\tilde{\delta}/2\pi \propto 75~\text{kHz}$ in a Yb-Ba crystal, and reduces it to $\tilde{\delta}/2\pi \propto 3~\text{kHz}$ in a Ba-Yb crystal.}
\label{fig:rotate_e_field}
\end{figure}

We can increase the size of the XCOM/ZSTR crossing in Fig.~\ref{sec:gen_all_mode_cooling} by applying an additional $\propto xy$ rotation shim to the crystal. This can significantly increase the coupling between these two modes. We plot this effect in Fig.~\ref{fig:rotate_e_field}, where we show that adding a radial shim term $V_{xy}= V_{DC,0}xy/2$ increases the size of the crossing to $\delta/2\pi \sim 75~\text{kHz}$ for a $^{171}\text{Yb}^{+}-^{137}\text{Ba}^{+}$ crystal. Unfortunately, if we use radial electric fields and rotation shims \textit{simultaneously}, the induced mode-mode coupling becomes ion-order dependent. This is because the uniform force from the radial electric field will add constructively/destructively with the differential force from the rotation shim in a way that result in a different total force being applied to either ion, breaking the symmetry of the problem for mixed-species crystals. We show this in Fig.~\ref{fig:rotate_e_field} where we show same crossing but for a $^{137}\text{Ba}^{+}-^{171}\text{Yb}^{+}$ crystal, showing the crossing narrows from $\delta/2\pi \sim 75~\text{kHz}$ to $\delta/2\pi \sim 3~\text{kHz}$ under this exchange. In state-of-the-art QCCD architectures \cite{moses_2023}, ion swaps such as this are treated similarly to ion losses, where the circuit is restarted as soon as the event is observed, so this order-dependence should not have an effect on its use in future architectures, but it could add difficulty to other types of experiments.

\section{Radial rotations}\label{sec:radial_rotations}

\begin{figure}[b]
\includegraphics[width=0.5\textwidth]{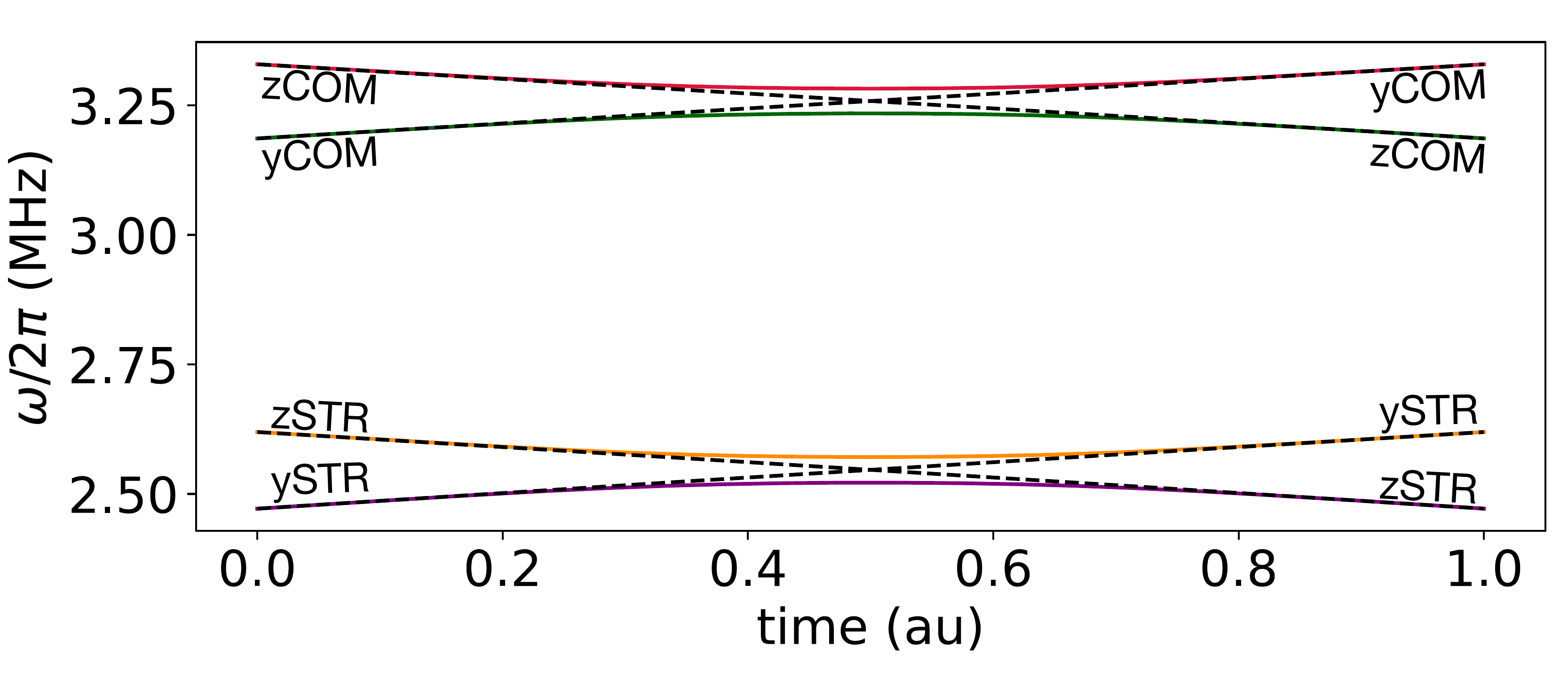}
\centering
\caption{Eigenfrequencies $\omega$ versus time (arbitrary units) for rotation of every radial mode with is corespondent pair from the other radial direction. We initialize the system such that the $y$-direction is more strongly confined than the $z$-direction, then we ramp on a $\propto zz-yy$ shim to cross each radial mode. We show this with (colored solid) and without (black dashed) a $\propto yz$ rotation shim of size $V_{r,0}$ added to couple the radial modes.}
\label{fig:rotate}
\end{figure}

The quadratic shims we have focused on so far rotate crystals between the axial and radial directions, i.e. shims that are $\propto xy$ or $\propto xz$. As discussed in Appendix~\ref{sec:trap_rotations}, because the ions are separated in $x$ each ion feels force that is equal and opposite its partner, causing the crystal to rotate in a manner that cancels the mode coupling from the shim that created it. Since the ions are assumed to be at rest on the RF null, i.e. at $y_{j}=z_{j}=0$ for each ion $j$, there is no such force created by $\propto yz$ rotation shims, and no such cancellation effect. It is, therefore, possible to couple the two radial directions of the trap with this type of shim, which then allows us to drive phrap transitions if we cross the modes with it on. Other than our applied DC shims, however, both radial directions are symmetric. Therefore, adjusting our electrodes to bring YCOM/ZCOM into degeneracy, necessarily brings YSTR/ZSTR into degeneracy, simultaneously. We show this in Fig.~\ref{fig:rotate}, where we initialize a crystal such that its $y$-modes are lower frequency than its $z$-modes. We then ramp on a $\propto yy-zz$ shim until the modes cross, while applying a radial rotation shim according to $V_{rad}=V_{DC,0}\sin^{2}(\pi t/t_{r})yz$ for the phrap duration $t_{r}$. Here, we can see that both mode pairs cross at the same time. We can straightforwardly interpret this as a rotation of one radial direction into other, where we initialize the crystal with one weak axis and one strong axis then rotate the weak axis 90 degrees, exchanging the weak and strong axis as well as the motion associated with each. 

\clearpage 

\bibliography{biblio}

\end{document}